\documentclass[sigconf,sort]{acmart}

\usepackage{xcolor}
\usepackage{cleveref}
\usepackage{enumitem}
\usepackage{float}
\usepackage{graphicx}

\AtBeginDocument{%
  }

\setcopyright{acmlicensed}

\acmISBN{978-1-4503-XXXX-X/2025/07}

\begin{document}

\newcommand{\yang}[1]{{\bf \color{red} [[Yang says ``#1'']]}}
\newcommand{\yi}[1]{{\bf \color{blue} [[Yi says ``#1'']]}}
\newcommand{\yuting}[1]{{\bf \color{orange} [[Yuting says ``#1'']]}}
\ifdefined\nocomment
\renewcommand{\yang}[1]{\iffalse{#1}\fi}
\renewcommand{\yi}[1]{\iffalse{#1}\fi}
\renewcommand{\yuting}[1]{\iffalse{#1}\fi}
\fi

\title{Next-User Retrieval: Enhancing Cold-Start Recommendations via Generative Next-User Modeling}

\author{Yu-Ting Lan}
\authornote{Both authors contributed equally to this research.}
\affiliation{%
  \institution{ByteDance}
  \city{Shanghai}
  \country{China}
}
\email{lanyuting.8807@bytedance.com}

\author{Yang Huo}
\authornotemark[1]
\authornote{Corresponding author.}
\affiliation{
  \institution{ByteDance}
  \city{Shanghai}
  \country{China}
}
\email{huoyang.sunny@bytedance.com}

\author{Yi Shen}
\affiliation{%
  \institution{ByteDance}
  \city{Shanghai}
  \country{China}
}
\email{shenyi.01@bytedance.com}

\author{Xiao Yang}
\affiliation{%
  \institution{ByteDance}
  \city{Beijing}
  \country{China}
}
\email{wuqi.shaw@bytedance.com}

\author{Zuotao Liu}
\affiliation{%
  \institution{ByteDance}
   \city{Shanghai}
  \country{China}
}
\email{michael.liu@bytedance.com}

\begin{abstract}
The item cold-start problem is critical for online recommendation systems, as the success of this phase determines whether high-quality new items can transition to popular ones, receive essential feedback to inspire creators, and thus lead to the long-term retention of creators. However, modern recommendation systems still struggle to address item cold-start challenges due to the heavy reliance on item and historical interactions, which are non-trivial for cold-start items lacking sufficient exposure and feedback. Lookalike algorithms provide a promising solution by extending feedback for new items based on lookalike users. Traditional lookalike algorithms face such limitations: (1) failing to effectively model the lookalike users and further improve recommendations with the existing rule- or model-based methods; and (2) struggling to utilize the interaction signals and incorporate diverse features in modern recommendation systems.

Inspired by lookalike algorithms, we propose Next-User Retrieval, a novel framework for enhancing cold-start recommendations via generative next-user modeling. Specifically, we employ a transformer-based model to capture the unidirectional relationships among recently interacted users and utilize these sequences to generate the next potential user who is most likely to interact with the item. The additional item features are also integrated as prefix prompt embeddings to assist the next-user generation. The effectiveness of Next-User Retrieval is evaluated through both offline experiments and online A/B tests. Our method achieves significant improvements with increases of 0.0142\% in daily active users and +0.1144\% in publications in Douyin, showcasing its practical applicability and scalability.

\end{abstract}


\maketitle

\section{Introduction}
The cold-start problem is one of the long-standing challenges in online recommendation systems when introducing new items, onboarding new users, or dealing with inherently sparse interaction data.
Specifically, tackling the item cold-start challenges in real-world short video recommendation systems such as Douyin \cite{yan2024trinity}, Kuaishou \cite{liu2024kuaiformer}, and Youtube \cite{covington2016deep} determines whether high-quality new items can transition to popular ones to bring benefits to the brand of the platform, receive essential feedback to inspire creators, and finally lead to long-term retention of creators.
Although tremendous research has been conducted on cold-start problems \cite{zhang2025cold}, industrial recommendation systems still struggle with item cold-start issues due to their heavy reliance on item ID and historical interactions, which are ineffective for new items with limited exposure and feedback, ultimately reducing the fairness, interaction and diversity of recommendations.

One promising approach to address item cold-start problems is lookalike modeling \cite{mangalampalli2011feature,zhu2021learning,peng2023finding}, a machine learning technique that identifies groups of users with similar behavioral and demographic traits to a seed user and enables the item recommendation based on similarity. These methods are especially effective for items at the cold-start stage, as they rely on prior knowledge of the seed users rather than ID features and historical interactions. However, traditional lookalike methods face two main issues. Firstly, existing rule- \cite{shen2015effective} or model- \cite{liu2019real} based approaches often directly compare all possible pairs between seed and available users to find lookalike users, failing to accurately model user relationships and improve recommendations \cite{rahman2024graph}. Secondly, they struggle to utilize interaction signals and incorporate diverse features of modern recommendation systems, which limits their integration into existing systems and their applicability across various scenarios.

Inspired by lookalike modeling and recent progress in generative recommendation \cite{rajput2023recommender, zhai2024actions,liu2024kuaiformer}, we propose a novel generative approach called Next-User Retrieval to enhance item cold-start recommendations in Douyin's short-video platform. Our research investigates three key challenges with Next-User Retrieval in Douyin as follows:

\begin{itemize}[leftmargin=2mm]
    \item \textbf{How does Next-User Retrieval draw insights from lookalike modeling to enhance item cold-start recommendations?} We should first define the seed user and then design a method to accurately model the similarity between the seed user and lookalike users. 
    
    \item \textbf{How can we generate the next user and integrate this modeling into industrial recommendation systems?} Douyin’s recommendation system involves billions of videos per day, resulting in an immense pool of candidates. Unlike traditional NLP tasks where models typically operate with a limited token set, this vast pool makes it computationally infeasible to calculate probabilities for all candidates using a naive softmax approach.

    \item \textbf{How do we optimize the performance of the transformer, especially for the item cold-start scenario?} One primary issue is that the cold-start scenario in Douyin still contains items of multiple stages, e.g., with no interactions,  limited interactions, and relatively sufficient interactions. Recommend items relying only on the recent lookalike users cannot handle the efficiency of all the stages of cold-start items. Additionally, the feature of lookalike users is constrained by latency and storage, and its feature domain is far smaller than the real requesting users with fully contextual features, thus requiring our dedicated designs.

\end{itemize}

To address the aforementioned challenges, we claim our contributions as follows:

\begin{itemize}[leftmargin=2mm]
    \item We define sequential users with the same sparse positive interactions (e.g., likes and comments) as lookalike users and develop a real-time exposure-grained feature system to process it. The feature of sequential users introduces dynamic interacted user information to strengthen the weak ID representations of cold-start items, and we utilize it to generate the next potential user who is most likely to interact with the given cold-start item.
    \item We introduce a transformer-based model with three dedicated designed losses to model the next-user generation. Unlike the naive softmax approach in NLP tasks, a contrastive loss is proposed to transform the discrete item prediction into item representation learning. In addition, a cross-entropy loss is introduced to utilize the samples of exposure but without interaction. An auxiliary loss is used to enhance sequential user representation learning. Those designs enable Next-User Retrieval to seamlessly integrate into Douyin’s HNSW-based retrieval streaming system \cite{malkov2018efficient}.
    
    \item We propose some effective tricks designed for the item cold start as follows: First, the unidirectional relationship of sequential users is introduced to the transformer with causal attention. Second, to address the limitations of relying only on sequence information, we enhance sequential user modeling by integrating additional item features, such as ID and category, as prefix prompt embeddings to assist the next-user generation. Third, a learnable [CLS] token is introduced to bridge the feature domain gap between sequential users and real requesting users. 
\end{itemize}

 We evaluate the effectiveness of our approach through offline experiments and online A/B tests. Our method achieves significant improvements, with increases of +0.0142\% in daily active users and +0.1144\% in publications in Douyin, a platform with over 600 million daily active users. Next-User Retrieval has been successfully integrated into Douyin's main recommendation system, showcasing its practical applicability and scalability.

\begin{figure*}[h]
  \centering
  \includegraphics[width=\linewidth]{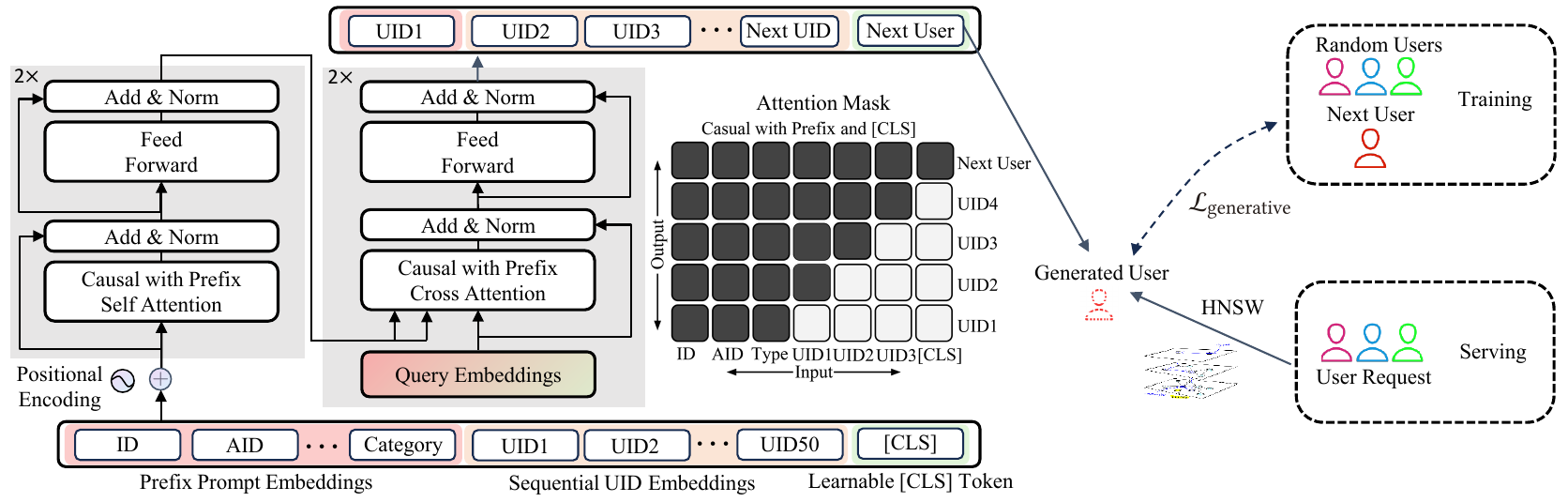}
  \caption{The framework of Next-User Retrieval. The dark and light cells in the attention mask indicate whether the token is allowed or not allowed to attend to the corresponding attention combination, respectively.}
  \label{model}
\end{figure*}

\section{Method}
This section describes how Next-User Retrieval works in Douyin's recommendation system. Firstly, we specify the problem definition and the pipeline of training and inference. Then, we present the feature engineering of interacted sequential users. Finally, the details of Next-User Retrieval framework and the loss function are given.

\subsection{Problem Statement}
In this section, we formulate the retrieval task as a next-user generation task, where the historical interaction sequence (such as likes and comments) of users is used to model the video’s generalized appeal and to generate the next user most likely to interact with it.

As demonstrated in \Cref{model}, in the training procedure, since we know the real next user $u$, we aim to maximize the following likelihood to model the most likely interacted user. 
\begin{equation}
\small
\operatorname{argmax} P\left(u, f_{u}) \mid \text{model}\left(\{(u_1), \ldots, (u_j), \ldots, (u_n)\}, (i, f_i)\right)\right).
\end{equation}
where $P(\cdot) \in \mathbb{R}^{|\mathcal{U}|}$ is the probability distribution over all users. $u_j$ denote the ID embedding of the $j^{th}$ user (UID) in the chronological order. $f_i$ and $f_u$ denotes the features of the given item $i$ and the real next user $u$, respectively.

During inference, the goal is to retrieve the cold-start items based on the similarity between the requesting user’s embedding and the generated next-user embeddings of each item. Specifically, when a user requests the recommendation system, the system computes the dot product between the user’s embedding and all generated next-user embeddings with HNSW, an Approximate Nearest Neighbor (ANN) based algorithm.

\subsection{Feature Engineering}
The recommendation system in Douyin needs to process
massive traffic requests in one second, which requires the feature system
to respond in real-time. The storage and latency constraints could be the
bottleneck of the Next-User Retrieval, and the traffic would increase linearly as the interacted user sequence's length and features increase. To achieve the constraints, we only store the user ID with positive interactions such as likes and comments and limit the maximum sequence length to 50.  


\subsection{Next-User Retrieval Pipeline}
We present the design of our Next-User Retrieval pipeline in this section. Next-User Retrieval takes the prefix prompt embeddings, sequential UID embeddings, and a learnable [CLS] token as input. Then, a dedicatedly designed causal attention with prefix prompts and [CLS] token in transformer is proposed to model the unidirectional relationship and optimize the next-user generation in cold-start scenarios. Then, three loss functions are combined to guide model training, enhancing both the generative ability and its robustness in cold-start scenarios.

\noindent \textbf{Sequential UID Embeddings:}
Since the interaction metrics such as likes and comments are displayed to content creators within Douyin's user interface, they can be considered essential feedback for encouraging long-term creator retention. We filter the interaction sequences with at least one of the aforementioned positive interactions. The filtered UID is stored in chronological order and mapped to a corresponding sequential UID embedding.

\noindent \textbf{Prefix Prompt Embeddings:} To address the limitations of relying only on sequential information, we enhance sequential UID modeling by using prefix prompt embeddings. Using prefix prompts with various ID and category features, items with no interaction can generate the first user. Items with relatively sufficient interactions can utilize the ID to maintain long-term interaction memory. This approach helps to improve the performance of multistage cold-start items with different exposures and interactions.

\noindent \textbf{Learnable [CLS] token:} Since we only store the interacted sequential UID embeddings, while the real requesting user has fully contextual features. To encode the feature domain prior knowledge to the transformer model, a learnable [CLS] token is appended to the end of sequential embedding, dynamically guiding the model to determine the appropriate generation mode for the sequential UID embedding or the next-user embedding for retrieval.



\noindent \textbf{Causal Attention with Prefix Prompts and [CLS] token:} As shown in \Cref{model}, to incorporate the inductive bias for unidirectional sequence relationships, we introduce the causal attention mechanisms into our transformer model, i.e., the generation of each token is conditioned on preceding tokens. This unidirectional relationship is motivated by the observation that users typically engage in actions such as commenting only after encountering some interesting comments, and that there exists a unidirectional relationship.

We slightly modify the causal attention for prefix prompts and [CLS] token as follows: Firstly, we modify the causal attention mechanism by leaving the prefix prompts unmasked. This enables the prefix prompt to learn an informative contextual representation. Then, the generation of the first UID is conditioned on the prefix prompts, and subsequent UIDs are generated sequentially conditioned on both the prefix prompts and all the preceding UIDs.
When generating the next user, the learnable [CLS] token is unmasked to switch the generation mode from the sequential UID to the user embedding. 

\noindent \textbf{Transformer-based Encoder-Decoder Architecture:} We employ a transformer-based encoder-decoder architecture for next-user generative modeling. The transformer encoder operates as follows:
\begin{equation}
\small
o_1^p, \dots, o_k^p, o_1^u, \dots, o_n^u, o_1^{[CLS]} = \text{Encoder}(p_1, \dots, p_k, u_1, \dots, u_n, [CLS])),
\end{equation}
where $p_i \in \mathbb{R}^d$ and $u_i \in \mathbb{R}^d$ denote the prefix prompts embedding and the sequential UID embedding, respectively. A learnable [CLS] token $[CLS] \in \mathbb{R}^d$ is appended to the end of the sequential input. $o_i^k \in \mathbb{R}^d$ denotes the encoder output of the $i^{th}$ token with type $k$.

The transformer decoder aggregates the sequential information from encoder outputs as follows:
\begin{equation}
\small
\hat{u}_1, \hat{u}_2, \dots, \hat{u}_{n+1}, \hat{u}_\text{next} = \text{Decoder}(q, (o_1^p, \cdots, o_k^p, o_1^u, \dots, o_n^u, o_1^{[CLS]})),
\end{equation}
where $q \in \mathbb{R}^{{(n+2)} \times d}$ represents the learnable query embeddings. With the query, the generated UID embedding $\hat{u}_i \in \mathbb{R}^d$ is conditioned on prefix prmopts outputs $o_1^p, \cdots, o_k^p$ and sequential UID encoder outputs $o_1^u, \dots, o_{i-1}^u$; The generated next user embedding $\hat{u}_\text{next} $ is conditioned on all the encoder outputs.

\subsection{Loss Function}
To model the concept of the next-user generation and integrate such modeling into existing industrial recommendation systems, we propose a combined loss function for model supervision as follows:
\begin{equation}
\small
\label{eq:generative_loss}
\mathcal{L}_{\text{generative}} = \lambda_1\mathcal{L}_{\text{contrastive}} + \lambda_2\mathcal{L}_{\text{CE}} + \lambda_3\mathcal{L}_{\text{auxiliary}}.
\end{equation}
\textbf{Contrastive Loss:} The $\mathcal{L}_{\text{contrastive}}$ aims to increase the similarity between the generated next user embedding and ground-truth interacted user embeddings and decrease the similarity with randomly sampled user embeddings.
The $L_\text{contrastive}$ is formulated as:
\begin{equation}
\small
\mathcal{L}_{\text{contrastive}} = -\sum_{i: R_{u_i \hat{u}_i}=1} \log \frac{\exp \left(f(u_i, \hat{u}_i) / \tau\right)}{\exp \left(f(u_i, \hat{u}_i) / \tau\right) + \sum_{j \neq i} \exp \left(f(u_j, \hat{u}_i) / \tau\right)},
\end{equation}
where \(R_{u_i \hat{u}_i}=1\) indicates that user interacted with the given item in the $i^{th}$ sample. The $u_i$ and $\hat{u_i}$ denote the requesting user embedding and generated next-user embedding, respectively.  The $u_j$ denotes the requesting user embedding from the $j^{th}$ sample. $f(\cdot)$ represents the similarity function (using the dot product,  which is the same as HNSW), and $\tau$ is the temperature parameter. The contrastive loss aims to transform the discrete next-user prediction into similarity-based item representation learning and enable the Next-User Retrieval to seamlessly integrate into Douyin's retrieval based on HNSW.

\noindent \textbf{Cross-Entropy Loss:} Since the samples that are exposed but not interacted with are also collected in our online streaming system, it is impossible to know which user will interact next for these samples, so they cannot be used in $L_\text{contrastive}$. However, these samples can pass through Douyin's recommendation funnel (retrieval, pre-rank, rank, and rerank) and reach exposure, indicating they are relatively high-quality and informative compared to random negative samples. $L_\text{CE}$ is proposed to better leverage this information as follows:
\begin{equation}
\small
\mathcal{L}_{\text{CE}} = -\left(\sum_{i: R_{u_i \hat{u}_i}=1} \log \sigma(f(u_i, \hat{u}_i)) + \sum_{i: R_{u_i \hat{u}_i}=0} \log(1 - \sigma(f(u_i, \hat{u}_i)))\right),
\end{equation}
where \(\sigma(\cdot)\) is the sigmoid function, and \(R_{u_i \hat{u}_i}=0\) indicates samples with exposure but without the interaction. The loss allows the model to make effective use of both interacted and non-interacted data, thus improving the model performance.

\noindent \textbf{Auxiliary Loss:} 
$\mathcal{L}_{\text{auxiliary}}$ serves as a data augmentation method to supervise the generation of the next UID and enhance the UID representation learning. The auxiliary loss is defined as:

\begin{equation}
\small
\mathcal{L}_{\text{auxiliary}} =\left(\sum_{i: R_{u_i \hat{u}_i}=1} ( \sum_{j=1}^{n+1} \| \text{sg}({u}_j) - \hat{u}_j \|^2 \right),
\end{equation}
where $u_j$ represents the $j^{th}$ ground-truth sequential UID embedding and $\hat{u}_j$ represents the generated UID embedding, respectively. The $\text{sg}(\cdot)$ denotes stopping gradient propagation to $u_j$ to avoid model collapse.

\section{Experiments and Results}
In this section, we present the offline and online experimental results with extended ablations. 
We list the following research questions (RQs) to guide the experimental discussion. \textbf{RQ1:} Does Next-User Retrieval achieve the best performance with the proposed modules?  \textbf{RQ2:} Can Next-User Retrieval bring improvement to the real-world recommendation system in Douyin?

\begin{table}[ht]
    \small
    \caption{Offline results of relative difference in Douyin's online learning system.}
    \centering
        \begin{tabular}{c|c|c|c}
        \hline ID & Variants & Recall@Top20 & Recall@Top50\\
        \hline
        0 & Next-User Retrieval&  0.4100 &  0.5859   \\
        \hline
        1 & Traditional Lookalike & -27.20\% & -22.11\% \\ 
        \hline
        2 & Mask Prefix Prompt & -14.15\% & -11.30\% \\
        \hline
        3 & Half the Sequence Length & -4.33\%  & -1.98\% \\
        \hline
        4  & w/o CLS Token & -0.03\% & -0.13\% \\
        \hline
        5 & w/o causal Attention &  -0.52\% & -0.22\% \\
        \hline
        \end{tabular}
    \label{tab:offline}
\end{table}

\noindent \textbf{Offline Results (RQ1):} We list Reall@topk and its relative difference as evaluation metrics. The experiments are conducted in Douyin's distributed training framework in an online learning way with hundreds of billions of training samples each day, and the metrics are recorded after the model is converged,. As illustrated in \Cref{tab:offline}, variant 1 takes interacted sequential UID embeddings as input and employs sum-pooling, as the traditional lookalike method. Variants 2 and 3 are conducted to evaluate the validity of the prefix prompt and the importance of the maximum length of sequential UIDs.  Variants 4 and 5 can be considered as the ablation for our modification of the transformer backbone. The offline ablation study analyzes the effectiveness of each module of our Next-User Retrieval to further answer \textbf{RQ1}.

\begin{table}[ht] 
\centering
\caption{Online A/B tests in Douyin. The symbol $-$ indicates the metric is not statistically significant.}
\small
\begin{tabular}{c|c|c|c}
\hline 
Variants & Daily Active Users & Publications & Interactions \\ \hline
Traditional Lookalike & $-$ & $-$ & -3.1800\% \\ \hline
Mask Prefix Prompt & +0.006\% & +0.0599\% & +2.3992\%  \\ \hline
Half the Sequence Length & +0.0083\% & +0.1144\% & +7.4535\% \\ \hline
Next-User Retrieval & +0.0142\% & +0.1144\% & +7.0515\% \\ \hline
\end{tabular}
\label{tab:online}
\end{table}

\noindent \textbf{Online Results (RQ2)}:
To measure the improvement of Next-User in our short-video recommendation system, we conduct a comprehensive online A/B test with 10\% users for at least one week. Since the ultimate goal is to inspire the long-term retention of creators, we take the daily active user as our key online metric. Notably, the improvement of 0.005\% in daily active users is a statistically significant change to give a satisfactory contribution since our platform has over 600 million active users per day. In addition, we take the publications of creators and the essential interactions to creators in the cold-start stages as proxy metrics. As shown in \Cref{tab:online}, Next-User Retrieval has demonstrated significant improvements in Douyin, which answers \textbf{RQ2}.

\section{Conclusion}
In this paper, we propose the Next-User Retrieval, a novel framework for enhancing cold-start recommendations via generative next-user modeling. Next-User Retrieval captures the unidirectional relationships among recently interacted users and utilizes these sequences to generate the next potential user who is most likely to interact
with the given cold-start item, which helps high-quality new items receive essential feedback to inspire
creators and thus leads to the long-term retention of creators. The effectiveness of Next-User Retrieval is evaluated through both offline experiments and online A/B tests. Our method achieves significant improvements with increases of 0.0142\% in daily active users and +0.1144\% in publications in Douyin. Next-User Retrieval has
been successfully integrated into Douyin’s main recommendation
system, demonstrating its practical applicability and scalability.

\newpage
\bibliographystyle{ACM-Reference-Format}
\bibliography{main}


\end{document}